\documentclass[a4paper,11pt]{article}
\usepackage{pos}

\title{String Geometry Theory and The String Vacuum}

\author*[a]{Matsuo Sato}

\affiliation[a]{Graduate School of Science and Technology, Hirosaki University\\ 
 Bunkyo-cho 3, Hirosaki, Aomori 036-8561, Japan}

\emailAdd{msato@hirosaki-u.ac.jp}

\abstract{String geometry theory is a candidate of the non-perturvative formulation of string theory. In this theory, strings constitute not only particles but also the space-time. In this review, we identify perturbative vacua, and derive the path-integrals of all order perturbative strings on the corresponding string backgrounds by considering the fluctuations around the vacua. On the other hand, the most dominant part of the path-integral of string geometry theory is the zeroth order part in the fluctuation of the action, which is obtained by substituting the perturbative vacua to the action. This part is identified with the potential for the string backgrounds and obtained explicitly. The global minimum of the potential is the string vacuum. The urgent problem is to find the global minimum. We introduce both analytical and numerical methods to solve it. }

\FullConference{Corfu Summer Institute 2023 "School and Workshops on Elementary Particle Physics and Gravity" (CORFU2023)\\
 23 April - 6 May , and 27 August - 1 October, 2023\\
Corfu, Greece\\}

\begin{document}
\maketitle

\section{Introduction}
\begin{figure}
\includegraphics[width=.3\textwidth]{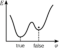}
\caption{A notion for a potential for the string backgrounds in a non-perturbative formulation of string theory}
\label{Potential}
\end{figure}

In string theory, there are extremely large numbers ($> 10^{500}$) of perturbatively stable vacua, which are called the string theory  landscape. Perturbative string theories cannot determine the true vacuum among them because they are defined only around local minima.  On the other hand, a non-perturbative string theory is thought to  determine the true vacuum (Fig.\ref{Potential}).  
In this review, we will derive a potential for string backgrounds from string geometry theory \cite{Sato:2017qhj, Sato:2019cno, Sato:2020szq, Sato:2022owj, Sato:2022brv, Honda:2020sbl, Honda:2021rcd, Sato:2023lls}, which is one of the candidates of non-perturbative string theory. 
The true vacuum can be determined by its minimum.
Main part of this proceeding is based on \cite{Nagasaki:2023fnz}.



\section{The idea of string geometry theory}
One of the fundamental problems in string theory is to determine the  six-dimensional internal space by a non-perturbative formulation of string theory. Thus, to consider what is the space-time in string theory is an important clue to understand what is a non-perturbative formulation of string theory. In perturbative string theories, the space-time is made of points, whereas a particle is made of a string. Because  it is thought that quantum space-time is fluctuated, natural generalization is that the space-time will be also made of strings in a non-perturbative formulation of string theory. This is the principle of string geometry theory. That is, not only particles but also the space-time are made of strings in string geometry theory.

\section{What we have done in string geometry theory}
String geometry theory is one of the candidates of a non-perturbative formulation of string theory. Evidences are as follows:
\begin{quote}
 \begin{itemize}
 \item 
 We can derive the path-integrals of the type IIA, IIB, SO(32) type I, and SO(32) and E8xE8 heterotic superstring theories from the single theory  by considering fluctuations around fixed backgrounds in the corresponding charts, respectively.
  \item 
The action is strongly constrained by T-symmetry in string geometry theory, which is a generalization of T-duality among perturbative vacua in string theory. 
  \item 
  The theory unifies particles and the space-time. That is, macroscopically,  the space-time = a string manifold,  and a particle = a fluctuation of a string manifold.
 \end{itemize}
\end{quote}

\section{A brief review on string geometry theory}
\begin{figure}
\includegraphics[width=.3\textwidth]{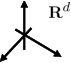}
\caption{A model space of finite dimensional manifolds
}
\label{FiniteModel}
\end{figure}

\begin{figure}
\includegraphics[width=.3\textwidth]{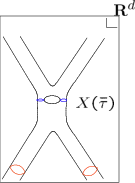}
\caption{Strings in $\mathbb R^{d}$.}
\label{Embedding}
\end{figure}

In the following, we consider only  the bosonic and closed sector of string geometry theory for simplicity. A model space of finite dimensional manifolds is just $\mathbb R^d$ as in Fig.\ref{FiniteModel}, whereas that of infinite dimensional manifolds has a non-trivial structure. 
The model space of string geometry has three kinds of coordinates:
\begin{quote}
 \begin{itemize}
 \item 
$\bar{\tau}$:  string geometry time $\in \mathbb R$
 \item 
 $\bar{h}$   : metric on a worldsheet $\Sigma$  ($\bar{h}$ is a discrete variable in the topology of string geometry.)
 \item 
 $X(\bar{\tau}): \Sigma|_{\bar{\tau}} \to \mathbb R^{d}$,
where the global time on $\Sigma$ is identified with string geometry time  $\bar{\tau}$, and  $\Sigma|_{\bar{\tau}}\cong S^1 \cup \cdots \cup S^1$ (Fig.\ref{Embedding})
 \end{itemize}
\end{quote} 
By considering any value of $\bar{\tau}$, any $\bar{h}$ and $X(\bar{\tau})$, a model space of string geometry, $E=\{ [\bar{\tau}, \bar{h}, X(\bar{\tau})] \}$ is obtained. 
String manifolds are constructed by patching open sets of the model space. 

\begin{figure}
\includegraphics[width=.3\textwidth]{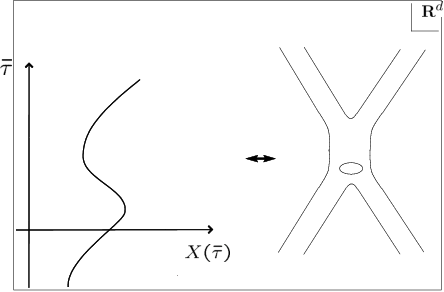}
\caption{Perturbative strings in string geometry}
\label{Moduli}
\end{figure}

Arbitrary two points with the same $\bar{\Sigma}$ in $E$ are connected continuously. 
Thus, there is a one-to-one correspondence between a Riemann surface in $\mathbb R^{d}$ and a curve parametrized by $\bar{\tau}$ from $\bar{\tau}\simeq -\infty$ to $\bar{\tau}\simeq \infty$ on $E$. 
That is, curves that represent asymptotic processes on $E$ reproduce the right moduli space of the Riemann surfaces in $\mathbb R^{d}$ as in Fig.\ref{Moduli}. 

The partition function of the bosonic and closed sector of string geometry theory is given by
\begin{eqnarray}
Z=\int \mathcal{D}G \mathcal{D}\Phi \mathcal{D}B e^{-S},
\nonumber
\end{eqnarray}
where the action is given by 
\begin{eqnarray}
&S =\int  \mathcal{D} h \mathcal{D} \bar{\tau} \mathcal{D} X(\bar{\tau})  \sqrt{G } e^{-2 \Phi} \left[ -R - 4 \nabla_{I} \Phi \nabla^{I} \Phi + \frac{1}{2} |H |^{2}  \right], \notag
\nonumber
\end{eqnarray}
where
\begin{quote}
 \begin{itemize}
 \item
$R$: scalar curvature of a metric $G_{IJ}$ on a string manifold,
 \item 
 $\Phi$:  scalar field  on a string manifold,
  \item 
 $H$:  3-form field strength of a 2-form $B$ on a string manifold.
 \end{itemize}
\end{quote} 
The index $I$ runs $d$ and $(\mu \bar{\sigma})$, and then, 
\begin{eqnarray}
\nabla_{I} \Phi \nabla^{I} \Phi
=
\nabla_{d} \Phi \nabla^{d} \Phi
+
\int d \bar{\sigma} \bar{e} \nabla_{(\mu \bar{\sigma})} \Phi \nabla^{(\mu \bar{\sigma})} \Phi,
\nonumber
\end{eqnarray}
where $(\mu \bar{\sigma})$ is an uncountably infinite dimensional index.

\section{Derive the path-integrals of perturbative strings 
in all the curved string backgrounds
from String Geometry Theory}
Let us consider fluctuations around backgrounds,
\begin{eqnarray}
G_{MN}=\bar{G}_{MN}+\tilde{h}_{MN}.
\nonumber
\end{eqnarray}
We fix the general covariance to the harmonic gauge,
\begin{eqnarray}
\bar{\nabla}^M \tilde{\psi}_{MN}=0,
\nonumber
\end{eqnarray}
where
\begin{eqnarray}
\tilde{\psi}_{MN}:=\tilde{h}_{MN}
-\frac{1}{2}\bar{G}^{IJ}
\tilde{h}_{IJ}\bar{G}_{MN}.
\nonumber
\end{eqnarray}
Because the degrees of freedom of strings are identified with $\tilde{\psi}_{dd}$ in \cite{Sato:2017qhj, Sato:2019cno, Sato:2020szq, Sato:2022owj, Sato:2022brv}, we set the other fluctuations zero. Actually we will derive the path-integrals of perturbative strings from $\tilde{\psi}_{dd}$. The background is fixed to so-called perturbative vacua \cite{Honda:2020sbl, Honda:2021rcd, Sato:2023lls},
\begin{subequations}\label{eq:sec3_condition2}
\begin{align}
\bar G_{dd} &= e^{2\phi[G,B,\Phi;X]}, \\
\bar G_{d(\mu\bar\sigma)} &= 0,\\
\bar G_{(\mu\bar\sigma)(\mu'\bar\sigma')}
&= G_{(\mu\bar\sigma)(\mu'\bar\sigma')}
= \frac{\bar e^3}{\sqrt{\bar h}}
G_{\mu\nu}(X(\bar\sigma))
\delta_{\bar\sigma\bar\sigma'},\\
\bar B_{d(\mu\bar\sigma)} &= 0,\\
\bar B_{(\mu\bar\sigma)(\mu'\bar\sigma')} 
&= B_{(\mu\bar\sigma)(\mu'\bar\sigma')} 
= \frac{\bar e^3}{\sqrt{\bar h}}\,
B_{\mu\nu}(X(\bar\sigma))
\delta_{\bar\sigma\bar\sigma'},\\
\bar\varPhi &= \varPhi
= \int d \bar\sigma\hat e\Phi(X(\bar\sigma)),
\end{align}
 \end{subequations}
 where $G_{\mu\nu}(x)$, $B_{\mu\nu}(x)$, and $\Phi(x)$ represent string backgrounds in the ten dimensions, and $\phi$ will be determined later. 

We normalize the leading part of the kinetic term by rescaling  $\tilde{\psi}_{dd}$ and delete the first order term by shifting  $\tilde{\psi}_{dd}$. This gives (condition 1), whose explicit form will be given later. 
Then, the action plus the gauge fixing term becomes
\begin{eqnarray}
S=S_0+
\int \mathcal{D} \bar{\tau}{\mathcal D}\bar{h} {\mathcal D}X(\bar{\tau})\tilde{\psi}_{dd}
H(-i\frac{1}{\bar{e}}\frac{\partial}{\partial X},-i\frac{\partial}{\partial\bar\tau}, 
X, \bar{h})
\tilde{\psi}_{dd},
\nonumber
\end{eqnarray}
where $S_0$ is the 0-th order terms and
\begin{align}
&H\Big(-i\frac1{\bar e}\frac{\partial}{\partial X},-i\frac{\partial}{\partial\bar\tau},X,\bar h\Big)\nonumber\\
&=\epsilon\Biggl( \frac12\int d\bar\sigma\sqrt{\bar h}
 G^{\mu\nu}\Big(-i\frac1{\bar e}\frac{\partial}{\partial X^{(\mu\bar\sigma)}}\Big)
  \Big(-i\frac1{\bar e}\frac{\partial}{\partial X^{(\nu\bar\sigma)}}\Big)
 - \frac12 e^{-2\phi}\Big(-i\frac{\partial}{\partial\bar\tau}\Big)^2\nonumber\\
&\quad
 + \int d\bar\sigma\bar e\Big(\bar n^{\bar\sigma}\partial_{\bar\sigma}X^{(\mu\bar\sigma)}
  + i\frac{\sqrt{\bar h}}{\bar e^2}G^{\mu\nu} \partial_{\bar\sigma}X^{(\rho\bar\sigma)} B_{\rho\nu}\Big)
 \Big(-i\frac1{\bar e}\frac{\partial}{\partial X^{(\mu\bar\sigma)}}\Big)
 +U,
\end{align}
where $U$ represents non-derivative terms and the ADM decomposition 
\begin{eqnarray}
\bar{h}_{mn}=
\left(
\begin{array}{cc}
\bar{n}^2+ \bar{n}_{\bar{\sigma}} \bar{n}^{\bar{\sigma}} & \bar{n}_{\bar{\sigma}} \\
\bar{n}_{\bar{\sigma}} & \bar{e}^2
\end{array}
\right)
\nonumber
\end{eqnarray}
is utilized. 

The differential equation for the propagator 
\begin{eqnarray}
\Delta_F\big(\bar h, X(\bar\tau), \bar\tau; \bar h', X'(\bar\tau'), \bar\tau' \big)
 = \big<\psi_{dd}''\big(\bar h, X(\bar\tau), \bar\tau),\psi_{dd}''(\bar h', X'(\bar\tau'), \bar\tau')\big>
\nonumber
\end{eqnarray}
is given by 
\begin{eqnarray}
H\Big(-i\frac1{\bar e}\nabla, -i\frac{\partial}{\partial\bar\tau}, X(\bar\tau), \bar h\Big)
\Delta_F\big(\bar h, X(\bar\tau),\bar\tau; \bar h', X'(\bar\tau'),\bar\tau'\big)
= \delta(\bar h - \bar h')\delta(X(\bar\tau) - X'(\bar\tau'))
 \delta(\bar\tau-\bar\tau').
\label{PropagatorDiffEq}
\end{eqnarray}

In order to compare with perturbative strings, 
   we take the Schwinger representation of the propagator by using the first quantization formalism, where 
$(\hat{\bar{h}}, \hat{X}, \hat{\bar{\tau}})$ are operators, $(\hat{p}_{\bar{h}}, \hat{p}_{X}, \hat{p}_{\bar{\tau}})$ are their conjugate momenta, and $|\bar{h}, X, \bar{\tau}>$ and $|p_{\bar{h}}, p_{X}, p_{\bar{\tau}}>$ are their eigen states.
Because (\ref{PropagatorDiffEq}) implies that the propagator is an inverse of $H$, it is given by matrix elements of $H$ with respect to the eigenstates, 
\begin{eqnarray}
\Delta_F(\bar h, X(\bar\tau), \bar\tau; \bar h', X'(\bar\tau'), \bar\tau')
&=& \big<\bar h, X(\bar\tau), \bar\tau|H^{-1}\big(\hat p_X(\bar\tau),p_{\bar\tau}, \hat X(\bar\tau), \hat{\bar h}\big)
  |\bar h', X'(\bar\tau'), \bar\tau'\big>\nonumber \\
&=&
\int _0^\infty dT\big<\bar h, X(\bar\tau), \bar\tau |e^{-T\hat H} |\bar h', X'(\bar\tau'), \bar\tau'\big>.
\nonumber
\end{eqnarray}

Because an observable must be invariant under diffeomorphism transformation, we consider 2-point correlation functions of diffeomorphism  invariant states (We integrate the boundary values in the end.),  
\begin{eqnarray}
\Delta_F(X_f; X_i | h_f; h_i) 
\coloneqq \int_0^\infty dT \big<X_f | h_f ; h_i\big\Vert_{\rm out}  
 e^{-T\hat H}\big\Vert X_i | h_f ; h_i\big>_{\rm in}, 
\label{time evolution}
\end{eqnarray}
where
\begin{eqnarray}
 \big\Vert X_i | h_f ; h_i\big>_{\rm in} 
 &\coloneqq \int_{h_i}^{h_f} \mathcal Dh'\big|\bar h', X_i, \bar\tau' = -\infty\big> \nonumber\\
\big< X_f | h_f ; h_i\big\Vert_{\rm out}
 &\coloneqq \int_{h_i}^{h_f} \mathcal Dh\big<\bar h, X_f, \bar\tau = \infty\big|.
\nonumber
\end{eqnarray}
$\Delta_F(X_f; X_i | h_f; h_i) $ can be written in a path integral representation because it is a time evolution of the states (\ref{time evolution}),
\begin{eqnarray}
&&\Delta_F(X_f; X_i | h_f; h_i) \nonumber \\ 
&=&
 \int_{h_i X_i, -\infty}^{h_f, X_f, \infty}  \mathcal{D} h \mathcal{D} X(\bar{\tau}) \mathcal{D}\bar{\tau} 
\int \mathcal{D} T  
\int 
\mathcal{D} p_T
\mathcal{D}p_{X} (\bar{\tau})
\mathcal{D}p_{\bar{\tau}}
\nonumber \\
&&
\exp \Biggl(- \int_{-\infty}^{\infty} dt \Bigr(
-i p_{T}(t) \frac{d}{dt} T(t) 
-i p_{\bar{\tau}}(t)\frac{d}{dt}\bar{\tau}(t)
-i p_{X}(\bar{\tau}, t)
\cdot \frac{d}{dt} X(\bar{\tau}, t)
+T(t) H(p_{\bar{\tau}}(t), p_{X}(\bar{\tau}, t), X(\bar{\tau}, t), \bar{h})\Bigr) \Biggr).
\nonumber
\end{eqnarray}
We move onto the Lagrange formalism from the canonical formalism by integrating out     ${p_X}_\mu $,
\begin{eqnarray}
{p_X}_\mu 
&= \frac{\bar e}{\sqrt{\bar h}}G_{\mu\nu}
 \Big(\frac1T\frac{dX^\nu}{dt} - T\bar n^{\bar\sigma}\partial_{\bar\sigma}X^\nu\Big) - i\frac1{\bar e}\partial_{\bar\sigma}X^\nu B_{\nu\mu}. 
\nonumber
\end{eqnarray}
As a result, we obtain
\begin{eqnarray}
\Delta_F(X_f; X_i | h_f; h_i) 
=
Z
\int_{h_i}^{h_f} 
\int_{X_i}^{X_f} 
\mathcal{D} h  \mathcal{D} X
e^{-S_{s}},
\label{pathint}
\end{eqnarray}
where 
\begin{eqnarray}
S_{s}
&=&
\int_{-\infty}^{\infty} d\tau \int d\sigma \sqrt{h(\sigma, \tau)} \frac{1}{2}\Biggl(\Bigl( 
h^{mn} (\sigma, \tau)
G_{\mu \nu}(X(\sigma, \tau))  
+i \epsilon^{mn} (\sigma, \tau)
B_{\mu \nu}(X(\sigma, \tau)) \Bigr) \partial_m X^{\mu}(\sigma, \tau) \partial_n X^{\nu}(\sigma, \tau)  \nonumber \\
&&+\alpha'\,R_{\bar{h}}\,\Phi(X(\sigma, \tau)) \Biggr),
\nonumber
\end{eqnarray}
where we have chosen perturbative vacua $\phi$ (condition 2).
(\ref{pathint}) is the path-integrals of all order perturbative strings in general backgrounds.

\section{Potential for string backgrounds
from string geometry theory}
Next, we will derive a potential for string backgrounds
from string geometry theory. In the last section, we have derived the path-integrals of perturbative string theories on the string backgrounds from the fluctuations around the perturbative vacua that include the backgrounds.
By setting the fluctuations 0,  the action becomes a classical action $S_0$, which can be obtained by substituting the perturbative vacua to the original action. The potential for string vacua  $V$  is given by  $V=-S_0$  because  $S_0$ is independent of the string geometry time $\bar{\tau}$.

For simplicity,  we take a particle limit,  $X^{\mu}(\sigma, \tau) \to x^{\mu}$,  where       
 \begin{eqnarray}
\int \mathcal{D}X \to \frac{1}{2\kappa_{10}}\int d^{10}x
\sqrt{-G(x)}.
\nonumber
\end{eqnarray}     

The conditions for perturbative vacua, which we imposed are explicity given as follows. 

\begin{quote}
 \begin{itemize}
 \item(condition 1) The first order term in the action vanishes by a shift of the fluctuation,                                   
\begin{eqnarray}
\tilde{\psi}_{dd}
\to
\tilde{\psi}_{dd}+f,
\nonumber
\end{eqnarray}
which means that the background $\phi$ (corresponding to $\tilde{\psi}_{dd}$) is on-shell,
\begin{eqnarray}
\nabla^2f  = -e^{-\Phi+\phi/2}(\nabla^2\phi + (\partial\phi)^2).
\end{eqnarray}
\item(condition 2) We chose the background $\phi$  so as to give the path-integrals of pertubative strings, which are Weyl invariant,
\begin{eqnarray}
R - \frac12 |H|^2 - \frac12\nabla^2\phi + \frac{\alpha}{2}(\partial\phi)^2
 - 3\nabla^2\Phi + 11(\partial\Phi)^2
 + \frac{17}{2}\partial^{\mu}\Phi\partial_{\mu}\phi
=0.
\end{eqnarray}
 \end{itemize}
\end{quote} 
We have already completed to derive the perturbation theory because these solutions exist.

By making an $\epsilon$ expansion around the flat background:
$G_{\mu \nu}=\eta_{\mu \nu} + \epsilon \tilde{G}_{\mu \nu}$, 
$|H|^2 = \epsilon |\tilde{H}|^2$, 
and 
$\Phi=\epsilon \tilde{\Phi}$,
we solve the conditions up to the second order,
\begin{eqnarray}
\phi &=& \epsilon\phi_1 + \epsilon^2\phi_2 \nonumber \\
f &=& f_0 + \epsilon f_1 + \epsilon^2 f_2,
\nonumber
\end{eqnarray}
for simplicity, 
and substitute them into
\begin{eqnarray}
V
&= \int d^{10}x \sqrt{-G}\Big[-e^{-2\Phi+\phi}
 \Big(R - \frac12|H|^2 - 2\nabla^2\phi - 2(\partial\phi)^2
 + 4(\partial\Phi)^2\Big)
 + e^{-\Phi+\phi/2}(\nabla^2\phi + (\partial\phi)^2)f\Big].
\nonumber
\end{eqnarray}
As a result,  we obtain a potential for string backgrounds up to the second order:
\begin{eqnarray}
V
&=& \frac{1}{2\kappa_{10}^2}\int d^{10}x \sqrt{-G}\Big( (-\frac32 - f_0 )
 (|H|^2-2R) \nonumber \\
 && -(\frac{97}{2}+24f_0) (|H|^2-2R) 
  \frac{1}{2\kappa_{10}}\int d^{10}x' \sqrt{-G} G(x;x')(|H'|^2-2R')
  \nonumber \\
  &&-(511+254f_0)\phi(|H|^2-2R)
+(1360+682f_0)(\partial \phi)^2\Big),
 \nonumber
\end{eqnarray}
where
$f_0$ is an arbitrary constant and $G(x;x')$ is a Green function that satisfies $\Delta G(x;x')=-\delta(x-x')$.
This potential has a multi-local form, naturally appearing as an effective action of quantum gravity \cite{Coleman:1988tj, Asano:2012mn, Hamada:2014ofa}.

Minimizing the potential will choose one of the solutions of supergravities and D-brane effective actions, which are obtained as a result of the consistency of the fluctuations (Weyl invariance in the perturbation theory). Such solutions are time dependent in general. Therefore, string geometry theory has a non-perturbative effect that can determine a true string vacuum.

\section{Conclusion}
In string geometry theory, we have identified perturbative vacua in string theory, which include general string backgrounds. From fluctuations around these vacua, we have derived the path-integrals of perturbative strings on the string backgrounds up to any order. We have also obtained  differential equations that determine a potential for string backgrounds. We have solved the differential equations up to the second order and obtained a potential explicitly up to that order.

\section{Outlook}
One of the important problems is to determine the string vacuum. An analytical approach is to assume realistic Calabi-Yau manifolds, where a region of the vacua to search is restricted. More generally, a numerical approach is to discretize  the potential by Regge calculus. In these approaches, one can determine a 6D manifold, expectation values of fields and a D-brane conifuguration. By integrating the 6D internal space, one can compactify the 10D effective theory of string theory and determine a (3+1)D effective action. By the standard phenomenological analysis, one can make the first prediction in string theory.

\section{Acknowledgement}
I appreciate my collaborators on string geometry theory:
K. Hatakeyama,
M. Honda,
T. Masuda,
K. Nagasaki,
Y. Sugimoto,
M. Takeuchi,
G. Tanaka,
T. Tohshima,
and
K. Uzawa.
I would like to thank 
H. Kawai,
J. Nishimura,
T. Yoneya,
and especially 
A. Tsuchiya
for valuable discussions.
I would also like to thank 
K. Anagnostopoulos and
G. Zoupanos
for warm hospitality on the conference. 
This work is supported by Hirosaki University Priority Research Grant for Future Innovation.

\end{document}